# Carbon $^{14}$C and Tritium as possible background sources in XENON1T

Yu. Shitov [a] and E. Yakushev [a,1]

[a] *Laboratory of Nuclear Problems, JINR, Dubna, 141980, Russia*

*E-mail*: yakushev@jinr.ru.

ABSTRACT: In this work, carbon $^{14}$C and Tritium were considered as possible background sources in the XENON1T experiment. The simulation results show that if $^{14}$C is located in dust particles with a characteristic size of tens of micrometers, then its beta spectrum is softened and can contribute to the low-background part of the spectrum (up to 20 keV) of the XENON1T experiment. In addition, it has been shown that the tritium spectrum is also significantly distorted due to the threshold effect and the form in which tritium is living in xenon. Comparison of the simulation results with experimental data allowed us to estimate the activity level of $^{14}$C at about 1500 decays / t / year, which gives the level of organic impurities containing $^{14}$C at the level of $2\times10^{-13}$ g/g. In the case of tritium background, spectrum distortions are already caused by nanoparticles. Wherein the shape of the spectrum from tritium sitting on the surface of the dust fits better into the experimental data than the spectrum from pure tritium. In addition, since the dust absorbs part of the decays, the total amount of tritium in xenon can be several times greater than assuming a background from pure tritium in xenon (the factor strongly depends on the size of the dust particles).

---

[1] Corresponding author.

**Contents**

**1. Introduction**

The discovery of an excess of events in the low-energy region in the XENON1T experiment [1] aroused great interest from the scientific community. In a short time, dozens of theoretical works appeared that interpreted this excess as New Physics of various kinds. However, within the framework of conservative traditions, a detailed analysis and the exclusion of more traditional causes of the observed effect are necessary. For example, background from unaccounted radioactive sources. With unprecedentedly low level of the background achieved in the XENON1T, new sources may arise that were not previously considered important. In presented paper, we consider the possible contribution of carbon $^{14}$C to the observed effect. In addition, the shape of beta-spectrum of tritium (considered as a background source by the XENON1T collaboration) can be distorted if T is sitting on the dust particles.

The beta radioactive carbon isotope $^{14}$C is constantly formed in the atmosphere in the reaction of neutron capture from cosmic rays (and other sources) by nitrogen $^{14}$N($n$, $p$)$^{14}$C. Having a period half-life of 5700 years, $^{14}$C forms carbon dioxide, consumed by plants during photosynthesis. Thus, $^{14}$C is presented in all organics on Earth. The presence of carbonaceous contamination on almost any surfaces is well known from XPS, low energy electron spectroscopy with radioactive sources [2], SEM [3]. The layer of organic compounds $CH_x$ is usually few nm thin and is present on surfaces considered clean even under vacuum conditions. Therefore contamination because of residual organic materials in an instrument (i.e. presence of $^{14}$C) can be considered as almost unavoidable. As the example, the need to take into account the presence of $^{14}$C on the surface was reported by the EDELWEISS dark matter search experiment [4].

Thus, any organic impurities (dust), if getting into liquid xenon of the active part of the detector, bring a radioactive $^{14}$C with it. Its amount is small, but with the ultra-high sensitivity achieved in the XENON1T experiment, it can be noticeable. Another important factor is the shape and size of the microparticles containing $^{14}$C. In its pure form, the $^{14}$C spectrum is too hard ($Q_\beta$ = 160 keV) compared to the region of the effect observed by XENON1T. However, if $^{14}$C is **inside dust particles**, then the surrounding organic layers will act as a moderator, squeezing the initial beta spectrum to the low-energy region. In addition, a well-known fact is the decrease in the efficiency of registration of low-energy particles (keV scale) in liquid xenon due to the quenching effect of light production (figure 1) - the **threshold effect**. Therefore, low-energy particles (for example, δ-electrons) formed in the tracks of beta particles are not always recorded, further softening the recorded beta spectrum. Given these two circumstances, the background from $^{14}$C may turn out to be similar to that observed in the XENON1T experiment.

All of the above in terms of moderation of the beta spectrum applies to tritium T, which can also exist in the detector, attached to dust particles. Therefore, its observed spectrum may differ significantly from the pure beta spectrum.

In this work, we simulated the behavior of $^{14}$C and T sources in different configurations of the compounds, and also taking into account the particle detection efficiency according to the experimental data of the XENON1T detector (see figure 1). The following discusses the methodology and results of these simulations.

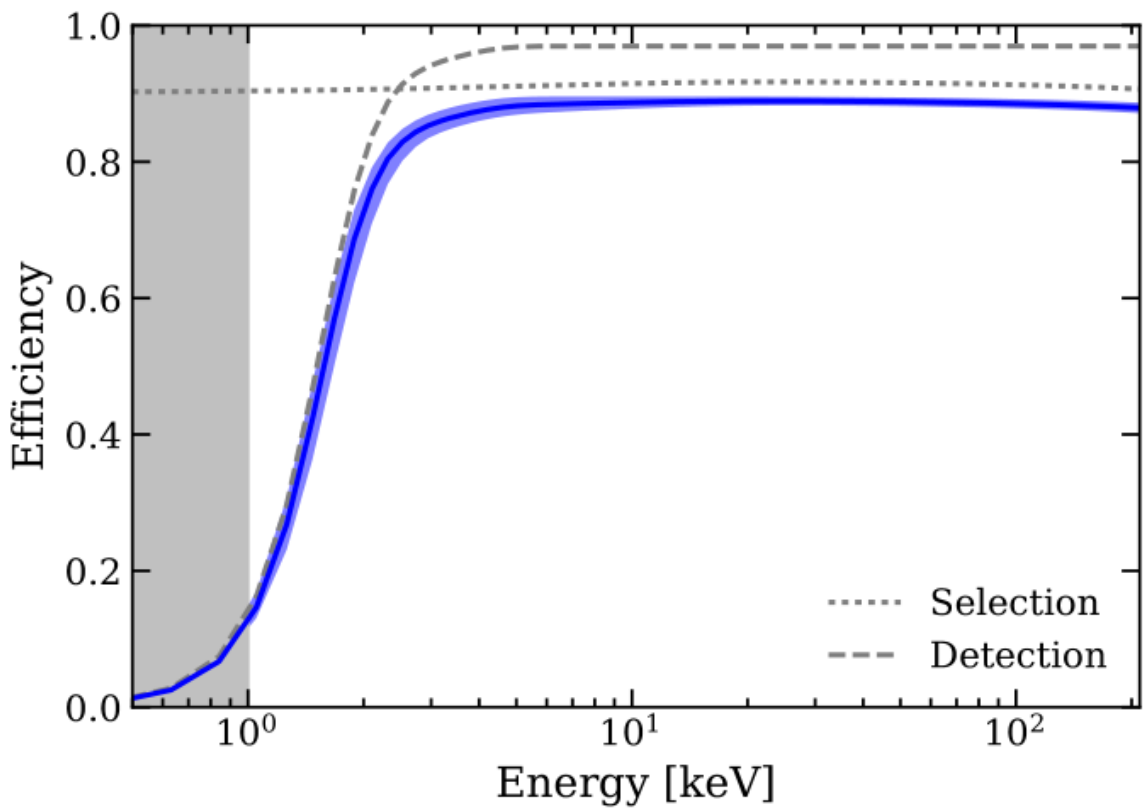

**Figure 1.** Signal detection (selection) efficiency (gray dashed line) and total recording efficiency (blue curve, blue band is ± 1 σ uncertainty) of the XENON1T detector as a function of particle energy. The gray bar is the threshold of the detector. The Source is figure 2 from [1].

## 2. Simulation technique

The configuration of the simulated geometry is shown in figure 2. Radioactive sources of $^{14}$C and T were simulated inside dust particles immersed in liquid xenon. It is hard to predict what kind of particles contains carbon. In this situation we assumed the most conservative approach: an organic dust. Our further conclusions will be even stronger for dust particles containing elements with higher Z and higher density, where energy losses per distance will increase significantly.

Polystyrene was chosen as the material of dust particles (standard GEANT4 material **G4_POLYSTYRENE** with density ρ = 1.06 g/cm$^3$, mass fractions C / H = 92.26 / 7.74%, respectively), as some characteristic type of particle.

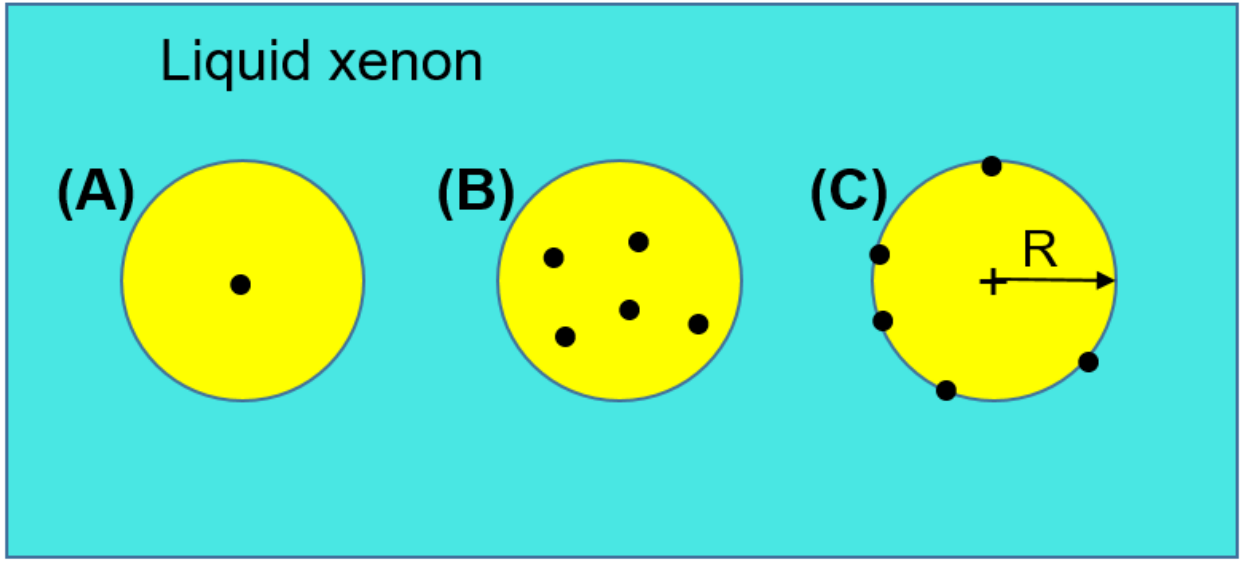

**Figure 2.** Radioactive sources (black dots) inside dust particles (yellow spheres) floating in liquid xenon. The radioactive nuclei were simulated in the center **(A)**, uniformly in volume **(B)** and on the surface **(C)** of dust particles.

**The radius R** of particles is a key parameter that determines the distortion of the initial spectrum; therefore, the task of the simulations was to select the particle radii of impurities that cause characteristic changes in the spectrum.

The second important parameter is the **location of the decaying nucleus on the particle**. Three configurations were played out, shown in figure 2: the radioactive nucleus is located: A) in the center of the dust particle, B) uniformly inside its volume, C) uniformly on its surface. Option A) is likely to be of purely academic interest and is provided for completeness. In reality, under real conditions, option B) likely to be realized for $^{14}$C, which should be randomly distributed over volume of the organic material, as we assumed the polystyrene particles. And option C) is expected for tritium, which can be planted and attached to the surface of dust particles.

In the simulations, the GEANT4.10.06 version [5] was used. Standard electromagnetic processes were simulated (directly taken from GEANT *TestEm4* example). Initial spectra of $^{14}$C and T were generated through built-in GPS mechanism of the GEANT4 package using calculated (scanned and digitized) spectra of $^{14}$C [6, figure 3] and T [7, figure 1]. As is known, special GEANT4 low energy physics classes for (*G4LowEnergy*, *G4LECS*, and *G4Penelope*) can be used for precision calculations, as well as the experimental spectra can be taken into account. However, for the purposes and tasks of this work, this is not required, since the mentioned corrections and effects are insignificant in comparison with the effects from the dust material, its size and the location of radioactive nuclei together with the threshold effect, which we are studying qualitatively. Each configuration (radii 1, 10, 50, 100, 150 μm for $^{14}$C, and 1, 100, 200, 400, 1000 nm for T) was simulated using $10^6$ events.

## 3. Results

The results of simulations for $^{14}$C are shown in figure 3, and simulations of the tritium are shown in figure 4 and the threshold effect was taken into account in both cases here as well as further in the simulations shown in figure 5. A number of interesting conclusions can be drawn from these figures, which we will discuss in detail in the next section.

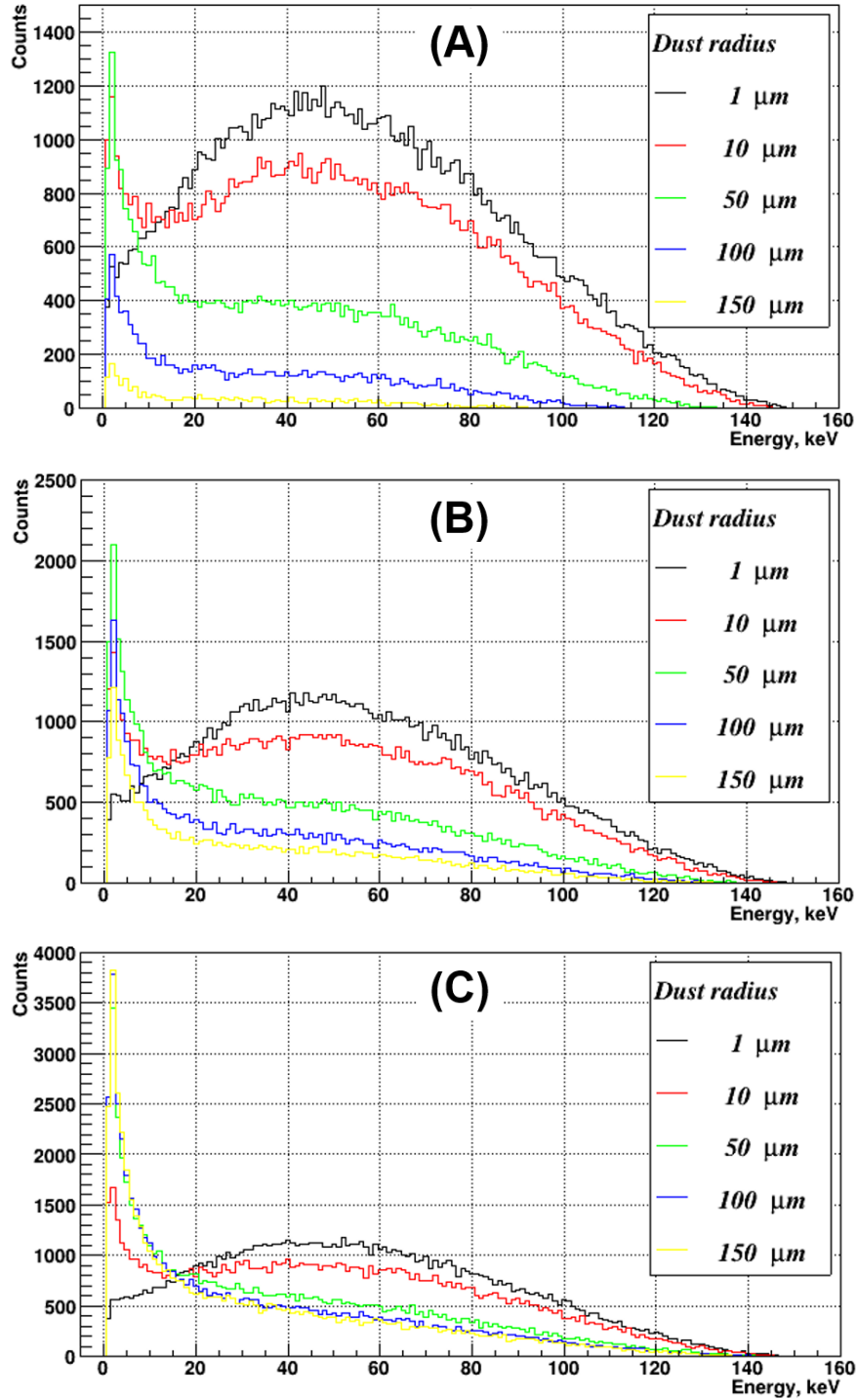

**Figure 3.** Spectrum of $^{14}$C with the isotope located in the center (**A**), uniformly in the volume (**B**), and uniformly over the surface (**C**) of dust particles of different radius (the correspondence of the particle radius to the line colors is given in the legend).

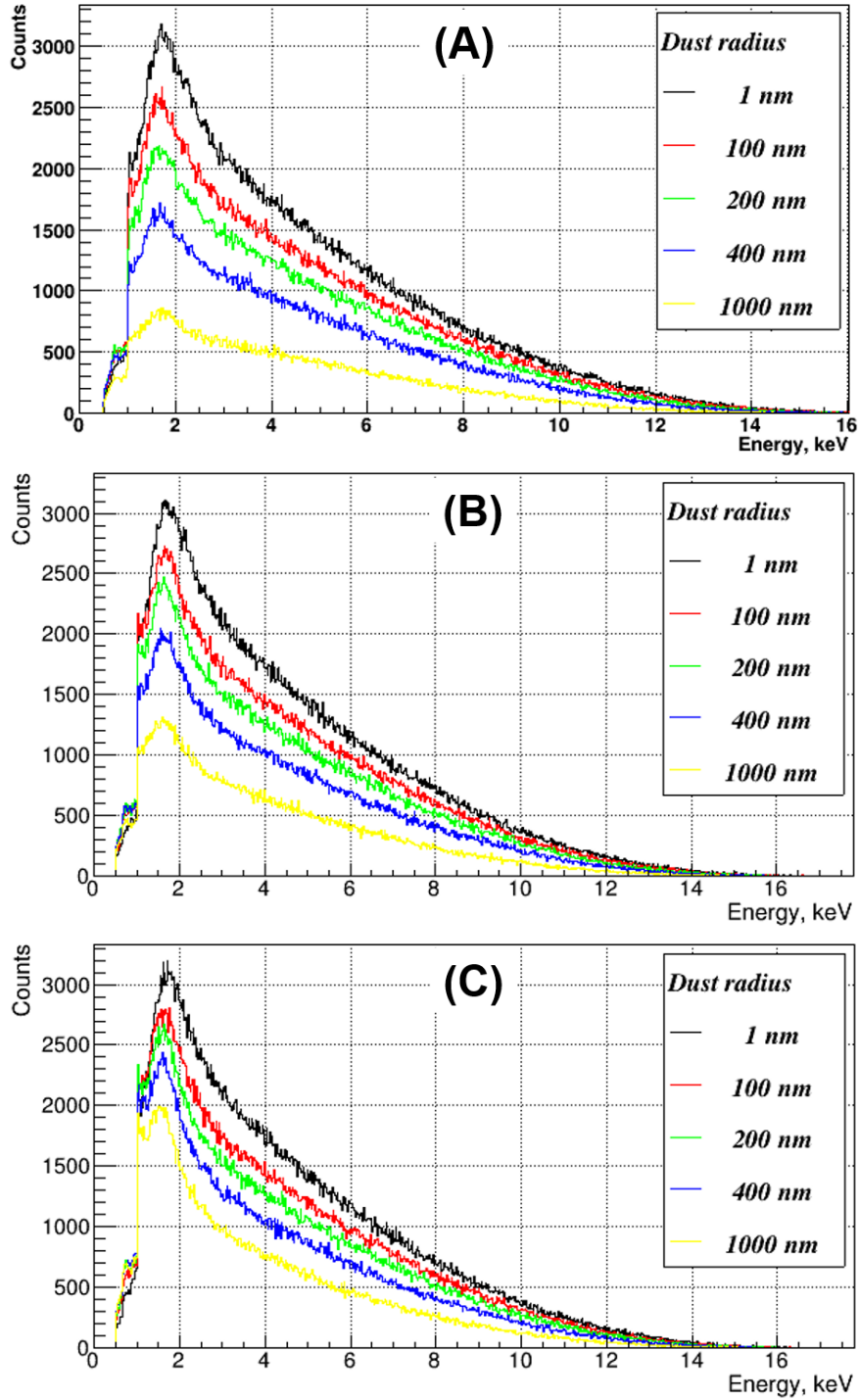

**Figure 4.** Spectrum of T at the location of the isotope in the center (A), uniformly in the volume (B), and uniformly over the surface (C) of dust particles of different radius (the correspondence of the particle radius to the line colors is given in the legend).

## 4. Discussion of the results

The simulations $^{14}$C and T allowed us to assess the potential contribution of these sources to the background of the XENON1T experiment. We discuss each of the isotopes separately.

*Simulations of $^{14}$C*

From the obtained graphs (figure 3) it is clearly seen that if $^{14}$C is sitting inside of dust particles, the beta spectrum softens and a peak arises in the 1-5 keV energy region. The intensity of this peak depends on the location of the radioactive nuclei (minimum in figure 3A and maximum in figure 3C). With a characteristic particle size of about 50 μm, the shape of the spectrum turns into a clear peak in the low-energy region up to 20 keV and a falling right tail. In this configuration, $^{14}$C could potentially be the background addition in the XENON1T energy spectrum. We emphasize that the most unpleasant option is to find the isotope on the surface of dust particles. Despite the fact that half of the beta particles fly directly into xenon, the second half sees the effective thickness of the dust particle is almost twice as large, and therefore gives a more intense 1-5 keV peak. The situation with tritium is similar, but the effect is less noticeable due to lower energy spectrum (see figure 5).

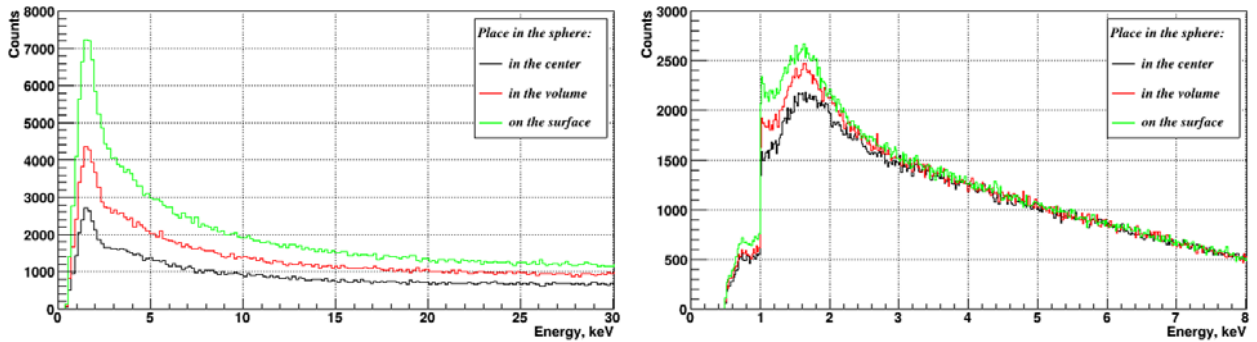

**Figure 5.** Spectra of $^{14}$C (left) and T (right) with the same characteristic particle size (50 μm and 200 nm for $^{14}$C and T, respectively), but different isotope locations according to figure 2.

*Tritium simulations*

Tritium has a substantially softer energy spectrum, therefore, a weakening of the registration efficiency of low-energy particles in the threshold region makes a significant contribution to the distortion of its shape (see figure 1, this will be discussed in the next section). The characteristic particle size that affects the spectrum is substantially smaller than at $^{14}$C - these are nanoparticles. Finally, the nature of the effect of the location of the isotope on the shape of the spectrum differs from $^{14}$C. In the case of tritium, the dust on which it sits works more as a shield. The larger the size, the lower the intensity, while the line shape itself changes less than for $^{14}$C (see figure 5 on the right). Although this change is also noticeable, what will be discussed further when comparing experimental data with simulated ones.

*Threshold effect*

As already discussed in the introduction, the threshold effect in the XENON1T detector (see figure 1) affects the shape distortion of the initial beta spectrum. The performed simulations show that this is especially noticeable on the softer tritium spectrum, where, even in the absence of any dust, a strong squeeze of the registered spectrum to the low-energy region is observed (figure 6 on the right). But even in the case of $^{14}$C, the effect is visible (figure 6 on the left).

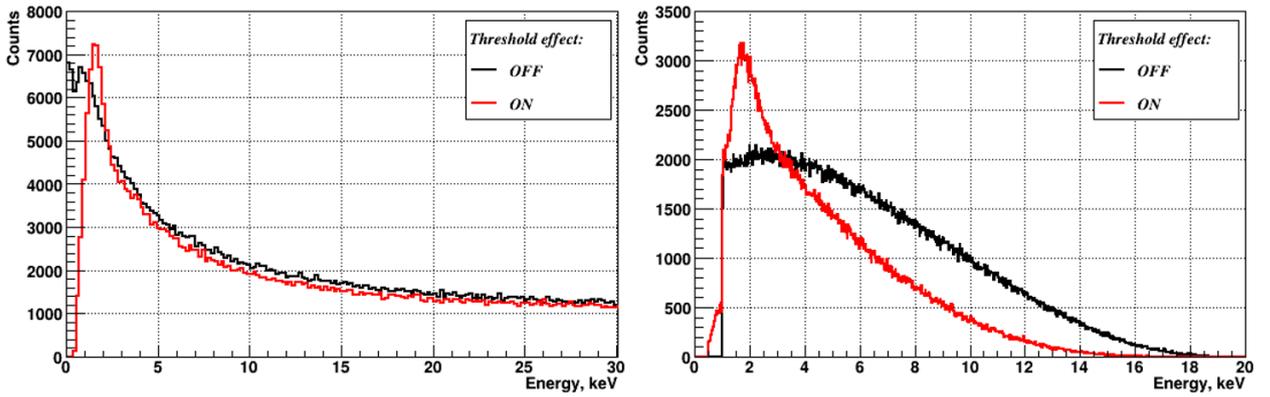

**Figure 6.** Spectra $^{14}$C (left) and T (right) with and without threshold effect. Simulation of $^{14}$C in a dust particle volume of 50 μm, tritium in its pure form without dust.

*Potential contribution to the XENON1T spectrum*

Realizing the behavior of $^{14}$C and T sources in a liquid xenon, we will further try to superimpose the obtained results on the experimental XENON1T spectrum. Having complete freedom in choosing parameters (we don't know the nature of the dust, nor the size, nor the location of the isotope), we cannot claim precision results. However, we can make rough, but quantitative estimates, which, in our opinion, will be correct in order of magnitude and therefore, in our opinion, are of interest in the aspect of the problem under discussion.

**The spectrum of $^{14}$C**, of course, in any configuration of the source location, in addition to a peak of interest in the 1-5 keV region, has a long descending high-energy tail in the region > 5 keV. Therefore, fit over the entire 1-30 keV spectrum naturally suppresses the low-energy peak. However, it is important to note that the estimate of the expected background (red curve in figure 7) in this region may be incorrect due to the neglect of $^{14}$C. Moreover, a decrease in the contribution of other isotopes to the 5-30 keV region would free up space for $^{14}$C tail. That is, it is not correct to fit $^{14}$C into the full 1-30 keV range, since correction and recalculation of the background model is required in the presence of $^{14}$C.

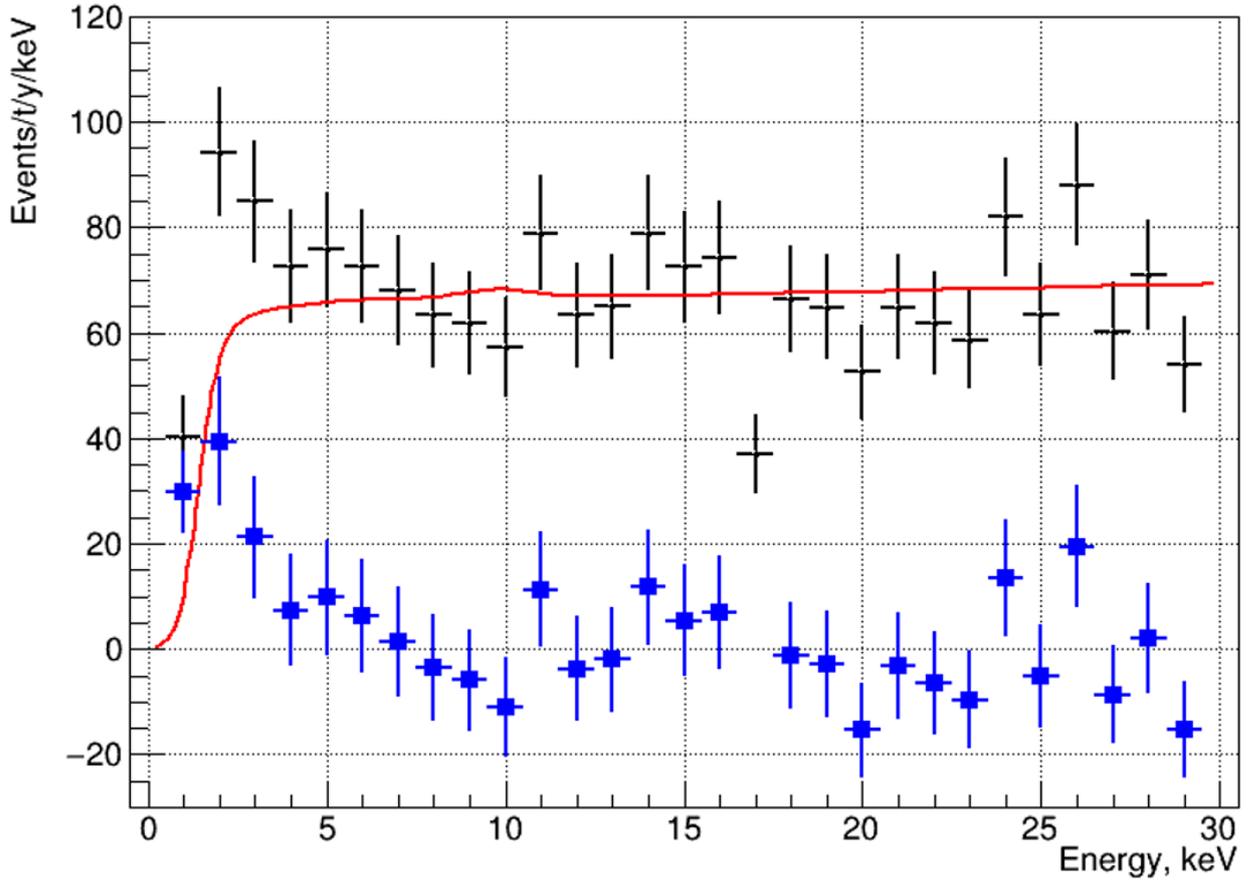

**Figure 7.** Spectrum of the XENON1T experiment (figure 4 from [1]). Black dots - experimental data, red curve - expected background, blue dots - experimental data with expected background subtracted.

Figure 8 shows the spectrum in the previously considered characteristic and most likely source location ($^{14}$C distributed randomly over the volume of spherical polystyrene dust particles with a radius of 50 μm), fitted into the XENON1T 1-5 keV data. The right tail is overestimated relative to the data, but, as explained above, tail correction can be done by rebalancing the background model. The resulting $^{14}$C activity is of the order of 1500 events / ton / year. Given the average activity of $^{14}$C at 238 Bq per kilogram of organics [8], the total mass of carbon dust in the plant is estimated at 0.2 μg / ton or 378 particles of the indicated size per ton of liquid xenon (or $2 \times 10^{-13}$ g / g). Unhappily even most sophisticated modern methods for detection of $^{14}$C, as accelerator mass spectrometry, are not able to reveal presence of $^{14}$C on the expected level. Such detection is especially hard as by our model the carbon is located in a single microparticle in about 2.7 kg of xenon. Detection of organic contamination on a level of $10^{-13}$ g / g seems to be even more complicated. Thus, only the XENON experiment itself is able to prove or reject the $^{14}$C and(or) T backgrounds, for example by observing a gradient versus depth in its activity. The presence of $^{14}$C and T can be also checked after accumulation of significantly more statistics, followed by a precise analysis of the spectrum shape. This is exactly what the collaboration envisages to do in the future.

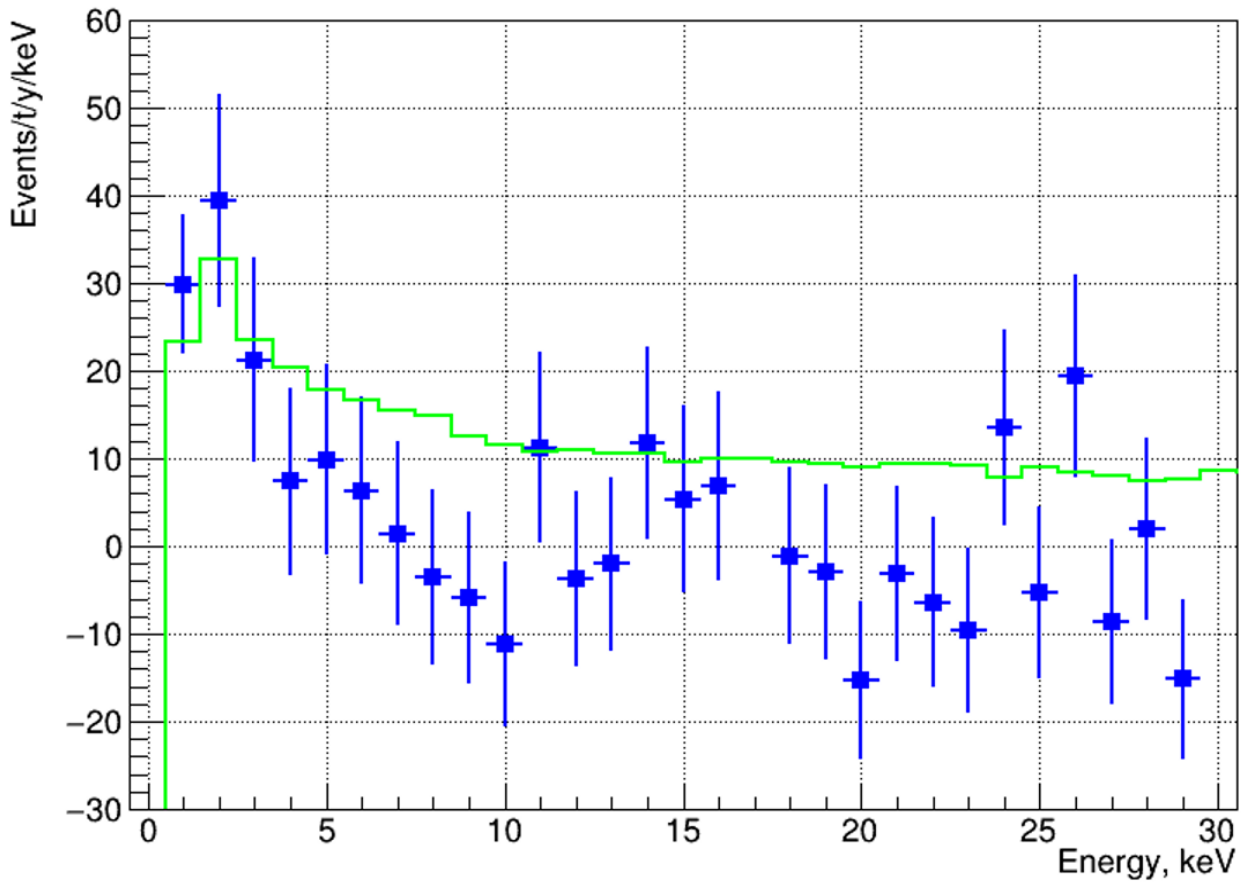

**Figure 8.** Contribution of $^{14}$C in the spectrum of XENON1T. The blue dots are the experimental data minus the expected background, the green histogram is $^{14}$C distributed randomly on spherical microparticles of polystyrene with a radius of 50 μm.

**The tritium spectrum** in the previously described characteristic and most likely source configuration (T distributed uniformly over the surface of spherical microparticles of polystyrene with a radius of 200 nm), fitted into the XENON1T experiment data, is shown in figure 9. Important point is: when tritium is on surface of dust particles, its total activity will be significantly higher than that calculated on the assumption that it is pure, since a significant part of the decays is not recorded. In our case, for the configuration under discussion, the real tritium activity is 538 decays / t / year (green histogram in figure 9), which is more than three times higher than the estimate of tritium activity 159 ± 51 decays / t / year obtained by the XENON1T collaboration in [1]. Note that for pure tritium, we got about 200 decays / t / year (red histogram in figure 9), which is consistent with the XENON1T estimation, and it is a good cross-check of the absence of critical errors in our calculation procedure. It can be seen also from figure 9 that the spectrum shape of tritium sitting on the dust, is more similar to the observed excess of events in XENON1T than the pure tritium spectrum ($\chi^2_{NORM}$= 1.2 versus $\chi^2_{NORM}$= 1.5, respectively). Thus, the presence of tritium in the form of compounds is more likely than in its pure form.

**Which background is most likely $^{14}$C or T?** We cannot quantitatively compare these hypotheses for two reasons. First, we do not know anything about the carrier (dust particles) of radioactive nuclei. And second, precision fitting of $^{14}$C spectrum requires rebalancing of the background model of the XENON1T experiment, the parameters of which are unknown to us. Our study shows that both isotopes and even their combination can be sources of excess events in the spectrum of the XENON1T detector - and this is the main qualitative conclusion. More accurate estimates and calculations can be performed further only by the XENON collaboration itself.

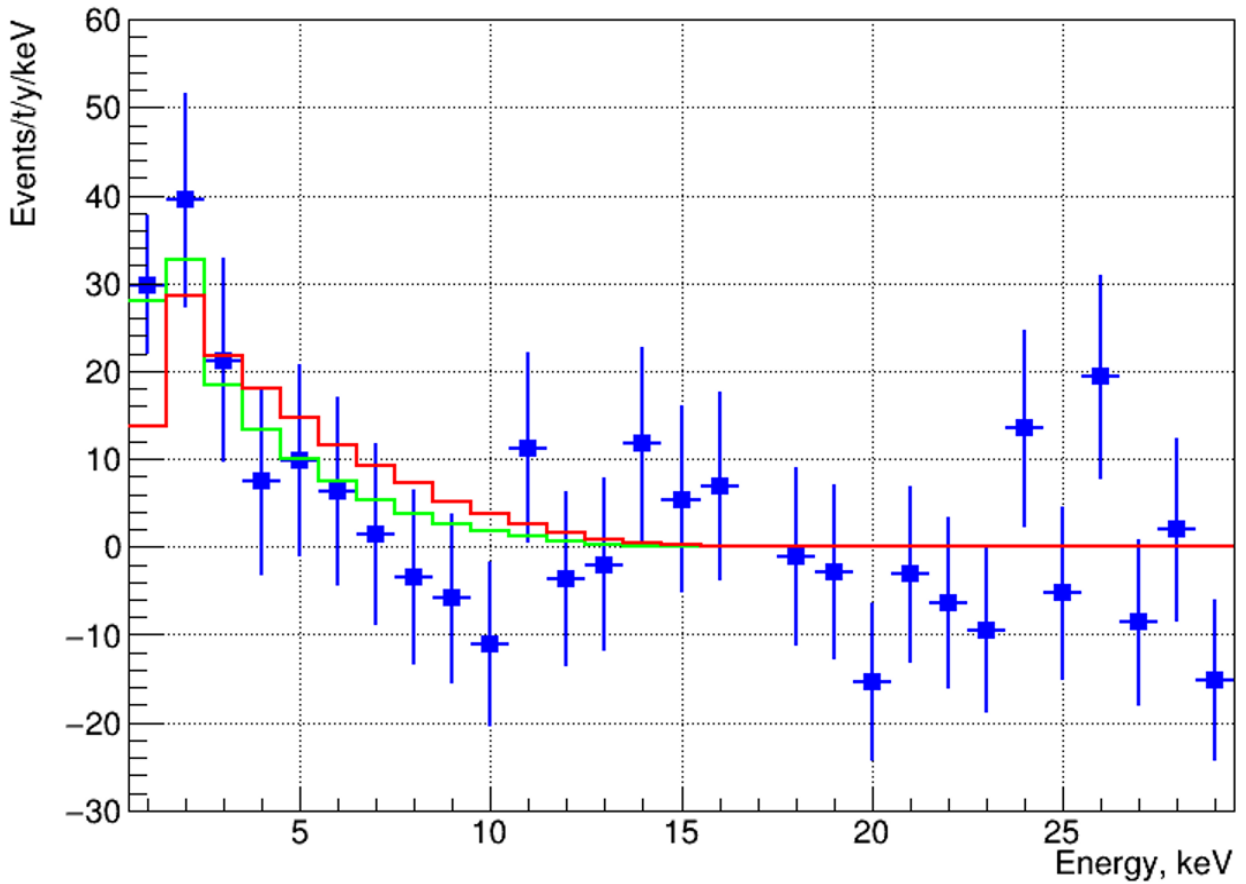

**Figure 9.** Contribution of tritium to the XENON1T spectrum. The blue dots are the experimental data without the expected background, the green histogram is T distributed uniformly over the surface of spherical polystyrene microparticles with a radius of 200 nm. The red histogram is the spectrum from pure tritium in liquid xenon. The threshold effect is taken into account in both cases.

**5. Conclusion**

The hypothesis of a possible contribution of radioactive $^{14}$C and T sources to the low-energy region of the XENON1T experiment was studied. Simulations of isotopes in liquid xenon were carried out under the assumption that they sit on some kind of organic, dusty particles, while spherical particles were simulated. Various options for the location of the sources were considered: in the center of the dust particle, uniformly in its volume (this is the most likely option for $^{14}$C) and the surface (which is more likely for tritium).

In addition, during the simulations, the threshold effect of the XENON1T detector was taken into account - a decrease in the detection efficiency near the threshold. The simulation results show that, with a characteristic particle size of tens of microns, the $^{14}$C spectrum is softened in such a way that it becomes similar to the spectrum observed in XENON1T. For tritium, because of the softer beta spectrum, a strong distortion of the initial form is introduced by threshold effect. At the same time, it becomes sensitive to dust particles of a much smaller characteristic size - at nm-scale. However, the resulting effect of dust on tritium differs from $^{14}$C. For tritium, dust works more as an absorber, reducing the intensity of the spectrum with a smaller change in its shape compared to $^{14}$C. Although at the same time, the change in shape occurs in a direction favorable for the description of the discussed effect. The shape of the spectrum from tritium sitting on the surface of the dust fits better into the experimental XENON1T data than the spectrum from pure tritium.

The fitting of the experimental data with the spectra we obtained made it possible to roughly estimate the level of $^{14}$C activity of about 1500 decays / t / year, which gives the level of organic impurities at the level of $2 \times 10^{-13}$ g / g. With respect to tritium, if it sits on dust, its real activity is significantly higher (in our case, more than 3 times) compared with the estimate assuming pure tritium as the background source. Simulations also show that the presence of tritium in the form of compounds is more likely than in its pure form.

Thus, according to the presented results, it will be interesting if the XENON1T collaboration will study the effects identified in this work for possible unaccounted $^{14}$C background and/or possible correction of tritium calculations.